\documentclass[conference]{IEEEtran}
\IEEEoverridecommandlockouts
\usepackage{cite}
\usepackage{amsmath,amssymb,amsfonts}
\usepackage{algorithm}
\usepackage{graphicx}
\usepackage{textcomp}
\usepackage{xcolor}
\usepackage{listings}
\usepackage{multirow}
\usepackage{subcaption}
\usepackage{balance}
\def\BibTeX{{\rm B\kern-.05em{\sc i\kern-.025em b}\kern-.08em
    T\kern-.1667em\lower.7ex\hbox{E}\kern-.125emX}}
\usepackage{url}
\PassOptionsToPackage{hyphens}{url}\usepackage{hyperref}

\newfloat{algorithm}{t}{lop}
\usepackage[noend]{algpseudocode}

\begin{document}

\title{Xorbits: Automating Operator Tiling for \\ Distributed Data Science}

\author{ 
\IEEEauthorblockN{
Weizheng Lu${}^1$, Kaisheng He${}^2$, Xuye Qin${}^{*2} $\thanks{${}^{*}$ Corresponding authors:  Yueguo Chen and Xuye Qin.}, Chengjie Li${}^2$, Zhong Wang${}^2$, Tao Yuan{}$^3$ \\ Xia Liao${}^4$, Feng Zhang${}^1$, Yueguo Chen${}^{*1}$\thanks{}, Xiaoyong Du${}^1$}
\textit{${}^1$Renmin University of China, ${}^2$Xorbits Inc.,}\\
\textit{${}^3$China Communications Information Technology Group, ${}^4$Tsinghua University}\\
luweizheng@ruc.edu.cn, \{hekaisheng, qinxuye, lichengjie, wangzhong\}@xprobe.io, \\ yuantao10@ccccltd.cn, liaoxia5018@163.com, \{fengzhang, chenyueguo, duyong\}@ruc.edu.cn, 
}

\makeatletter
\def\lst@makecaption{%
  \def\@captype{table}%
  \@makecaption
}
\makeatother

\lstset{
language=Python, 
basicstyle=\ttfamily\footnotesize, 
breaklines=true, 
keywordstyle=\bfseries\color{blue}, 
morekeywords={self}, 
emph={},
emphstyle=\bfseries\color{Rhodamine}, 
commentstyle=\itshape\color{black!50!white}, 
numbers=left,
numberstyle=\footnotesize, 
frame=single,
}

\maketitle
\begin{abstract}

Data science pipelines commonly utilize dataframe and array operations for tasks such as data preprocessing, analysis, and machine learning. The most popular tools for these tasks are pandas and NumPy. However, these tools are limited to executing on a single node, making them unsuitable for processing large-scale data.
Several systems have attempted to distribute data science applications to clusters while maintaining interfaces similar to single-node libraries, enabling data scientists to scale their workloads without significant effort. However, existing systems often struggle with processing large datasets due to Out-of-Memory (OOM) problems caused by poor data partitioning.
To overcome these challenges, we develop Xorbits, a high-performance, scalable data science framework specifically designed to distribute data science workloads across clusters while retaining familiar APIs. The key differentiator of Xorbits is its ability to dynamically switch between graph construction and graph execution.
Xorbits has been successfully deployed in production environments with up to 5k CPU cores. Its applications span various domains, including user behavior analysis and recommendation systems in the e-commerce sector, as well as credit assessment and risk management in the finance industry.
Users can easily scale their data science workloads by simply changing the import line of their pandas and NumPy code.
Our experiments demonstrate that Xorbits can effectively process very large datasets without encountering OOM or data-skewing problems. 
Over the fastest state-of-the-art solutions, Xorbits achieves an impressive 2.66$\times$ speedup on average. In terms of API coverage, Xorbits attains a compatibility rate of 96.7\%, surpassing the fastest framework by an impressive margin of 60 percentage points.
Xorbits is available at \url{https://github.com/xorbitsai/xorbits}.  

\end{abstract}

\begin{IEEEkeywords}
scalable data science, dataframe, array, tiling, computation graph
\end{IEEEkeywords}

\section{Introduction}

Data science (DS) pipelines are increasingly prevalent in today's world, owing to the emerging applications of machine learning (ML), artificial intelligence (AI), and data-driven business intelligence (BI)~\cite{yan2020AutoSuggest, petersohn2020Scalable}.
Dataframe and array systems, specifically pandas~\cite{mckinney2010Data} and ~NumPy\cite{harris2020Array}, form the most substantial portion of data science pipelines. The two libraries have gained widespread recognition among developers globally~\cite{stackoverflow2023Survey} primarily attributed to their various operators, flexible usage, and user-friendly interfaces. However, these two packages suffer scalability problems as they cannot distribute to multi-cores or multi-nodes, missing big data processing capability~\cite{GlobalInterpreterLock}. As datasets rapidly grow and workloads become more complex, there is an ever-increasing need to scale these libraries by utilizing computational resources far beyond what a single CPU-only node can offer. 
Several frameworks, such as PySpark~\cite{zaharia2016Apache}, Dask~\cite{rocklin2015Dask}, Ray~\cite{moritz2018Ray}, mpi4py~\cite{dalcin2005MPI}, and Modin~\cite{petersohn2021Flexible}, have been developed to address scalability concerns for data science workloads. However, our empirical studies reveal two significant issues with these tools. 
First, they struggle with handling extremely large datasets due to poor data partitioning and out-of-memory (OOM) problems, particularly in data-skewing scenarios. 
These systems build the computation graphs and partition data primarily by estimating the size of initial data sources ahead of runtime~\cite{petersohn2020Scalable}. 
However, the size of in-memory data can fluctuate after executing a series of data science operators, potentially diverging from the initial size. This can lead to memory exhaustion on specific worker nodes, especially during shuffle-intensive operations such as \texttt{groupby} or \texttt{merge}.
Second, they are not fully compatible with widely-used pandas and NumPy APIs~\cite{fromToPySparkPandas, daskDataFramesBestPractices}. This necessitates substantial code rewriting from users and is problematic as most data scientists only specialize in data modeling rather than parallel programming. These challenges significantly hinder users aiming to scale their data science workloads.

To address these problems, we develop Xorbits, a scalable data science engine that enables parallel execution of workloads like data-loading, preprocessing, scientific computing, analyzing, machine learning, etc. 
First, to optimize computation graphs at a finer granularity, we design three types of graphs—tileable graph (logical plan), chunk graph (coarse-grained physical plan), and subtask graph (fine-grained physical plan), in conjunction with a multi-stage \textit{map-combine-reduce} programming model.
Second, we introduce a novel dynamic tiling approach for automatic graph construction. This approach can switch between graph construction and graph execution, enabling us to build graphs by leveraging metadata from execution.
It considers the current operator's actual input data shape. Leveraging this real-time metadata, Xorbits can effectively partition and process data without encountering OOM issues, even when the data's shape substantially deviates from the initial data source.
Third, given the tiled graph, we employ graph-level and operator-level fusion. We also apply various optimizations, such as an intermediate storage service that utilizes different levels of storage devices, especially memory.

Xorbits can act as a drop-in replacement for the single-node data science libraries (i.e., pandas, NumPy, HuggingFace's datasets, etc) while ensuring competitive performance. 
Since Xorbits' APIs are identical to the original libraries, using Xorbits simply requires replacing the import code. As a result, users can scale data science programs with as many computing resources as they desire.
As of October 2023, Xorbits and its predecessor, Mars~\cite{mars}, have earned 3.4k stars on GitHub and have nearly 3k weekly downloads. Our users have successfully deployed Xorbits into their production environments, with up to 5,000 CPU cores. Xorbits has significantly expedited various data science and machine learning tasks such as financial risk management, fraud detection, user behavior analysis, and e-commerce recommendations.

We evaluate Xorbits' performance on dataframe and array operations using a combination of industry-standard benchmarks, such as TPCx-AI~\cite{brucke2023TPCxAI} and TPC-H~\cite{boncz2014TPCH}, along with real-world workloads. Compared to the fastest state-of-the-art solutions, Xorbits achieves a substantial 2.66$\times$ average speedup on data science operations that are supported by all baseline systems. Regarding API coverage, Xorbits attains a compatibility rate of 96.7\%, surpassing the fastest framework by an impressive margin of 60 percentage points. To provide an understanding of our optimizations, including dynamic tiling, we present an ablation analysis that delves into their effects on performance.

To summarize, this paper makes the following contributions:

\begin{itemize}
    \item We present the design of Xorbits, formalize three types of computation graphs, and propose the multi-stage \textit{map-combine-reduce} computation model.
    
    \item We propose the dynamic tiling approach, which leverages metadata from execution to partition data automatically. 
    
    \item We provide the details of optimization and implementations, including graph fusion, intermediate storage service, and auto re-chunking, .
    
    \item We evaluate Xorbits against existing systems like PySpark, Dask, and Modin on various benchmarks. We demonstrate significant speedups over these systems on data science pipelines, data analysis, and array computing workloads. Additionally, Xorbits supports a wide range of APIs and usage patterns.
    
\end{itemize}
\section{Background And Motivation}
\label{sec:background_motivation}

\subsection{Background}
\label{subsec:background}

The daily work of a data scientist primarily revolves around tasks such as loading, preprocessing, transforming, and conducting feature engineering on data~\cite{yan2020AutoSuggest, petersohn2020Scalable,POCLib}. To accomplish these tasks, they often utilize dataframe and array systems like pandas~\cite{mckinney2010Data} and NumPy~\cite{harris2020Array}.
According to the annual Stack Overflow Developer Survey, which gathered data from 67,231 respondents, the usage of NumPy and pandas accounts for 20.25\% and 18.97\% respectively among diverse programming languages and frameworks~\cite{stackoverflow2023Survey}. These percentages place them just behind \textit{.Net} and well ahead of other frameworks like \textit{React Native} and \textit{Apache Hadoop}.
NumPy, which underpins almost every Python scientific computing library, offers a n-dimensional array data type (\texttt{ndarray}) and array-aware functions. These functions enable operations such as matrix multiplication, finding the median of an array, indexing, etc.
Pandas, which supports a more user-friendly approach to modern data analysis than SQL, offers a functional interface that encourages quick and simple data exploration~\cite{petersohn2020Scalable}.
The main data structure of pandas is the dataframe (\texttt{DataFrame}), denoted as a tuple $(A, R, C, T)$. Here, $A$ represents an $m \times n$ array containing the data entries of the dataframe. $R$ is an array consisting of $m$ row labels, $C$ is an array containing $n$ column labels, and $T$ is an array that specifies the types for each column~\cite{petersohn2021Flexible}.
Dataframes support a wide range of operations, including both relational and non-relational ones. Relational operators, such as \texttt{join} (equivalent to the relational \texttt{JOIN}), enable combining data from different sources. Non-relational operations, like \texttt{pivot}, offer flexible data manipulation capabilities.

Both pandas and NumPy are in-memory, single-node, CPU-only computing engines.
As datasets continue to grow in size and exceed the memory capacity of a single computing node, the demand for a more advanced distributed solution is becoming increasingly urgent.
To solve this problem, the community has invested significant effort into building frameworks such as PySpark~\cite{zaharia2016Apache}, Dask~\cite{rocklin2015Dask}, Ray~\cite{moritz2018Ray}, Modin~\cite{petersohn2020Scalable},  and mpi4py~\cite{dalcin2005MPI}.
Due to the widespread popularity of pandas and NumPy, these frameworks have attempted to mimic their APIs, enabling users to transition to the new tools seamlessly.

\subsection{Observation}
\label{subsec:observation}

Our empirical study reveals that these frameworks exhibit poor scalability and migration issues for users. We conduct tests using the TPC-H benchmark\cite{boncz2014TPCH}, which comprises a total of 22 queries, to assess the scalability and usability of these systems. We use three scale factors (SFs): 10, 100, and 1000. Initially, we implemented a version using pandas and subsequently ported the code to other frameworks. Table~\ref{tab:failed_queries} presents the number of failed queries. Note that the PySpark version is developed using the pandas API on Spark (formerly known as the Koalas project) rather than Spark SQL. Based on our observations, these frameworks exhibit two primary issues.

\vspace{-0.5em}
\begin{table}[!ht]
\centering
\caption{Number of failed queries on TPC-H benchmark.}
\label{tab:failed_queries}
\begin{tabular}{c|cccc}
\hline
SF   & pandas & PySpark & Dask & Modin \\ \hline
10   & 0      & 3       & 1    & 0     \\ \hline
100  & 17     & 3       & 1    & 1     \\ \hline
1000 & 22     & 4       & 5    & 22    \\ \hline
\end{tabular}
\end{table}
\vspace{-0.5em}

\textbf{Scalability and Performance}.
These systems exhibit poor scalability and struggle to handle large datasets effectively. While they may be able to successfully execute queries with SF10, they encounter failures as the data size increases. To investigate the reasons behind these failures, we analyze their performance on the SF1000 dataset and summarize the findings in Table~\ref{tab:failed_reason}.
Modin on Ray offers better compatibility with pandas and can handle SF10 datasets. However, it faces challenges when dealing with larger data sizes, leading to OOM problems and termination of Ray workers. Running all 22 queries with Modin on SF1000 is difficult.
Similarly, Dask has five failed queries, two hanging queries, and three queries encountering OOM problems. This empirical study demonstrates the limited scalability of these systems, as they struggle with large datasets.

\vspace{-0.5em}
\begin{table}[!ht]
\centering
\caption{Reasons that frameworks fail on TPC-H SF1000.}
\label{tab:failed_reason}
\begin{tabular}{c|ccc}
\hline
Reason            & PySpark & Dask & Modin \\ \hline
API Compatibility & 3       & 0    & 0     \\ \hline
Hang              & 0       & 2    & 0     \\ \hline
OOM or Killed     & 1       & 3    & 22    \\ \hline
Total             & 4       & 5    & 22    \\ \hline
\end{tabular}
\end{table}
\vspace{-0.5em}

\textbf{API Compatibility}. 
These systems face API compatibility issues, and migrating data science workloads from a single machine to clusters is challenging. Spark, the most popular big data engine, fails mostly due to the absence of some pandas APIs. As a result, users frequently need to spend hours, or even days, searching for workarounds when encountering API issues. Notably, both Spark and Dask openly acknowledge that cannot achieve 100\% compatibility with pandas, as stated in their official documentation~\cite{fromToPySparkPandas, pysparkBestPractice, daskArrayBestPractices, daskDataFramesBestPractices}.
It requires users to have a deep understanding and practical experience with the frameworks they are using. Additionally, users often need to rewrite their single-node code to make their program able to run.
Dask offers both array and dataframe functionalities, but it's not easy to scale horizontally. When using Dask Array, users need to specify the chunk size. If the chunk size is too small, Dask will generate many task graph nodes (resulting in overhead). If the chunk size is too large, the data may not fit into memory~\cite{daskChoosing}. In some cases, incorrect chunk size configuration can prevent the program from running successfully. For instance, Dask's \texttt{qr} and \texttt{svd} functions only support specific matrix shapes, such as tall-and-skinny or short-and-fat matrices. If users fail to follow these chunking rules, Dask will throw exceptions.
Users must define chunk sizes explicitly using the \texttt{rechunk} function, as exemplified in the Dask Array example presented in Listing~\ref{lst:dask_chunk}. 
Since Dask DataFrame and pandas API on Spark only partition data on rows, they lack support for flexible operators. A prime example is Dask's inability to accommodate operations like \texttt{iloc}, which involve row slicing. Example of Dask DataFrame in Listing~\ref{lst:dask_chunk}, which is quite common in the data science community, will throw exceptions. Systems like Ray and mpi4py are general-purpose parallel computing engines. If users want to use Ray or mpi4py, they must build their distributed programs from scratch. In short, scaling the single-node data science code is not straightforward.

\subsection{Key Objectives}
\label{subsec:motivation}

Since we open-sourced our large-scale data framework, we have collected nearly 4,000 feedback from industry and academia. Based on these comments and recent progress in data science and artificial intelligence, we have distilled and identified the most pressing requirements of the data science community.

\begin{lstlisting}[
mathescape=true,
xleftmargin=2em,
xrightmargin=0.5em,
framexleftmargin=0.5em,
morekeywords={self,as},
caption=API compatibility issues of Dask.,
numbers=none,
label=lst:dask_chunk]
import dask.array as da
import dask.dataframe as dd

# Dask must specify chunk size
n = 10000
a = da.random.random(size=(n,n))
$\textbf{a = a.rechunk(chunks=(n, 1))}$
Q, R = da.linalg.qr(a=a)

# Dask failed with iloc
df = dd.read_parquet("<path>")
$\textbf{df = df.iloc[10]}$
\end{lstlisting}

\begin{itemize}
    \item \textbf{High Scalability and High-performance}. Data analytic and extract-transform-load (ETL) tasks in enterprises often need to process terabytes of data or beyond, so the data framework must be scalable enough to cope with the growing volume of data. 
    Furthermore, faster data processing speed can yield multiple benefits, such as quicker results for data analysts, more efficient utilization of computing resources, and cost savings.
    \item \textbf{API Compatibility}. API compatibility stands as another vital consideration.
    Most users start their data science journey by learning pandas and NumPy. In the industry, a common scenario involves conducting experiments on small datasets using a single node and then scaling these workloads onto clusters without the need to modify the code.
    \item \textbf{Python First}. As Python prevails in data science and artificial intelligence, users would like to use Python as its core or maybe the only programming language. They expect that Python could cover the whole data lifecycle, including data ingestion, preprocessing, model training, tuning, and inference. Since developers only need to be proficient in one language and the corresponding software stack, this reduces staff and employers' costs.
\end{itemize}

\textbf{Opportunity.}
None of the existing solutions mentioned before can meet these requirements. These points drive us to develop the Xorbits project, a high-performance data science engine that is capable of handling terabytes of data and beyond and is compatible with single-node libraries. When building the project, we face the dual challenges. First, the framework should scale seamlessly across a large number of computing nodes and can handle very large datasets. 
Data-skewing, one of the toughest issues that every big data framework faces, should be avoided. Second, the tiling of data should be done behind the scenes. Since single-node libraries do not have tiling or partitioning operations, data partitioning should be hidden to keep APIs compatible. 
\section{The Xorbits System}
\label{sec:xorbits_system}

In this section, we first introduce the architecture overview of Xorbits. Next, we present the user interface and the computation graphs.

\subsection{Overview}

\begin{figure}[ht]
    \centering
    \includegraphics[width=0.95\linewidth]{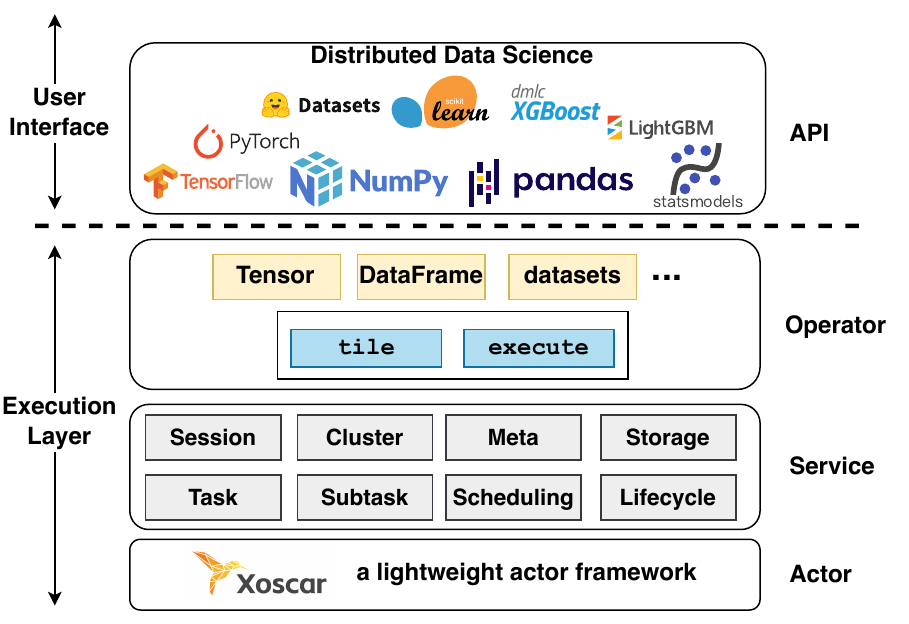}
    \caption{Overview of the Xorbits architecture.}
    \label{fig:architecture}
\end{figure}

\textbf{Architecture}. The architecture of Xorbits is depicted in Figure~\ref{fig:architecture}. Users can scale their data science workloads using the very familiar APIs of pandas and NumPy. Based on the distributed pandas and NumPy, Xorbits offers functionalities including data loading, preprocessing, and distributed machine learning, akin to PyTorch's dataloader~\cite{paszke2019PyTorch}, HuggingFace's datasets~\cite{huggingfaceDatasets}, XGBoost~\cite{chen2016XGBoost}, or scikit-learn~\cite{pedregosa2011Scikitlearn}. Internally, an API is defined as an operator. Our two most fundamental data structures are \texttt{Tensor} and \texttt{DataFrame}, where \texttt{Tensor} represents distributed arrays, and \texttt{DataFrame} denotes distributed dataframes. Xorbits implements several services to support the execution of scalable data science. 
Each service plays a particular management role. For instance, the session service creates, maintains, or destroys a session on a Xorbits cluster. All these services are based on an actor framework called Xoscar. Users can start a Xorbits cluster on any infrastructure like bare metal or Kubernetes. Note that this paper focuses on scaling data science workloads by dynamic tiling and operator fusion, so distributed machine learning and actor model of Xoscar are not main contributions of this paper.

\textbf{Xorbits Cluster}. When creating a Xorbits cluster, two kinds of daemons should be started: supervisor and worker. Users should choose a node (e.g., bare metal, container, virtual machine, etc.) to start the supervisor and spawn the workers on other nodes. The supervisor's main job is to manage subtasks, sessions, scheduling, etc. The worker is dedicated to task execution, carrying out the actual computational processes. Once the cluster is set up, users can submit their workloads to the supervisor, which will distribute tasks to workers.

\textbf{Novelty}.
First, we introduce the dynamic tiling approach (Section~\ref{sec:dynamic_tiling}) to automatically build data science computation graphs and tile data. Dynamic tiling can 1) prevent data-skewing and OOM issues, 2) generate optimized graphs, and 3) avoid re-partition code and keep APIs compatible. Second, we propose our novel graph fusion algorithm (Section~\ref{subsec:graph_optimization}) to fuse computation graphs for better performance. Third, we have performed substantial optimization work in areas such as scheduling~\ref{subsec:scheduling} and storage services~\ref{subsec:storage}. Fourth, to keep API compatible, we design auto rechunk algorithm (Section~\ref{subsec:auto-rechunking}).

\subsection{User Interface \& Application Scenarios}

\textbf{User Interface}. 
\texttt{DataFrame} and  \texttt{Tensor} are the two main distributed computing features provided by Xorbits.
Xorbits can act as a drop-in replacement for pandas and NumPy. The Xorbits' APIs (function signatures, arguments, and semantics) are exactly the same as those of the original packages. 
As demonstrated in Listing.~\ref{lst:api_example}, users can easily scale their data science workloads by changing lines of the import code and adding an \texttt{init} method to tell which runtime Xorbits should connect. When using Xorbits, users are relieved from manually specifying chunk sizes, the number of partitions, or performing operations like \texttt{repartition}. 
When migrating to Xorbits, users do not need extra effort to change their single-node code because Xorbits offers seamless integration and compatibility with the original packages. 
Xorbits can be effortlessly installed using the command \texttt{pip install xorbits} within a Python environment, eliminating the need for Java or any compilation processes.

\begin{lstlisting}[
mathescape=true,
xleftmargin=2em,
xrightmargin=0.5em,
framexleftmargin=0.5em,
morekeywords={self,as},
caption=Drop-in replacement of Xorbits.,
numbers=none,
label=lst:api_example]
import xorbits
import xorbits.numpy as np
import xorbits.pandas as pd
# init Xorbits runtime locally
# or connect to an Xorbits cluster
xorbits.init(http://<ip>:<port>)

# array example
a = np.random.rand(n,n)
Q, R = np.linalg.qr(a=a)
print(Q)

# dataframe example 1
df = pd.read_parquet("<path>")
df = df.groupby("A").agg("min")
print(df)

# dataframe example 2
df = pd.read_parquet("<path>")
filtered = df[df["col"] < 1]
print(filtered.iloc[10])
\end{lstlisting}

\textbf{Application Scenarios}.
As Xorbits provides distributed versions of dataframes and arrays with the very familiar APIs, it is a powerful tool for data scientists working on data-intensive applications. For example, there are three types of workloads in a real-world fraud detection workflow: dataframe-based ETL from raw logs, graph-based processing, and neural-network-based deep learning~\cite{yu2023Vineyard}. 
Xorbits can be adopted in workflows of ETL, analysis, and ML.
We have received feedback from our open-source community, attesting to the successful deployment of Xorbits into their production environments. Notably, the largest known Xorbits cluster has over five thousand CPU cores.
Our typical use cases include workflows for e-commerce recommendation systems, where data scientists analyze user behavior logs, and financial fraud detection. Moreover, Xorbits offers scalability for other DS libraries. For example, machine learning libraries like scikit-learn can be distributed with Xorbits' \texttt{Tensor} and \texttt{DataFrame}.

\subsection{Computation Graph}
\label{computation_graph}


Xorbits use directed acyclic graphs (DAGs) to describe the data dependency and the operator
execution order. 
There are three different types of computation graphs
in Xorbits: the tileable graph (logical plan), the chunk graph (coarse-grained physical plan), and the
subtask graph (fine-grained physical plan). 

\begin{figure}[ht]
    \centering
    \includegraphics[width=1\linewidth]{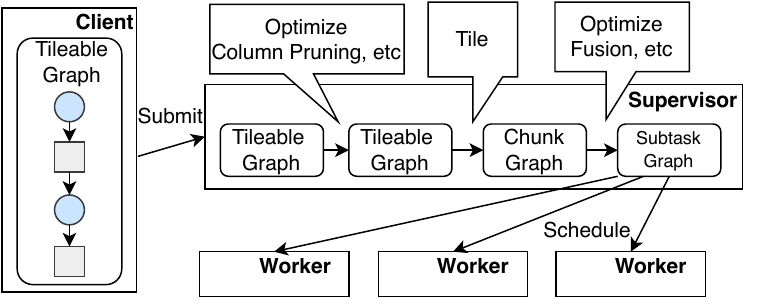}
    \caption{Workflow of optimizing and scheduling the computation graph.}
    \label{fig:workflow}
\end{figure}

\begin{figure*}[ht]
  \begin{subfigure}{0.23\textwidth}
    \centering
    \includegraphics[width=1\linewidth]{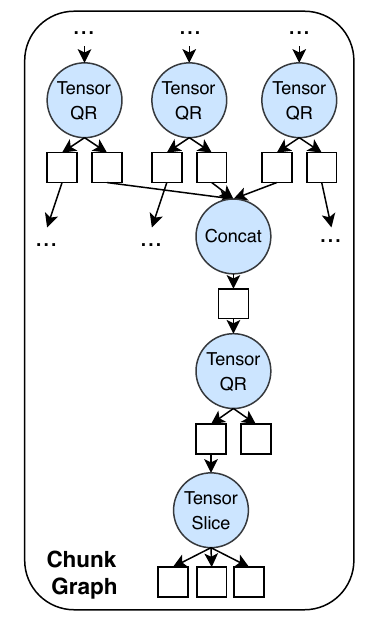}
    \caption{Array example.}
    \label{fig:array_graph}
  \end{subfigure}
  \begin{subfigure}{0.45\textwidth}
    \centering
    \includegraphics[width=0.96\linewidth]{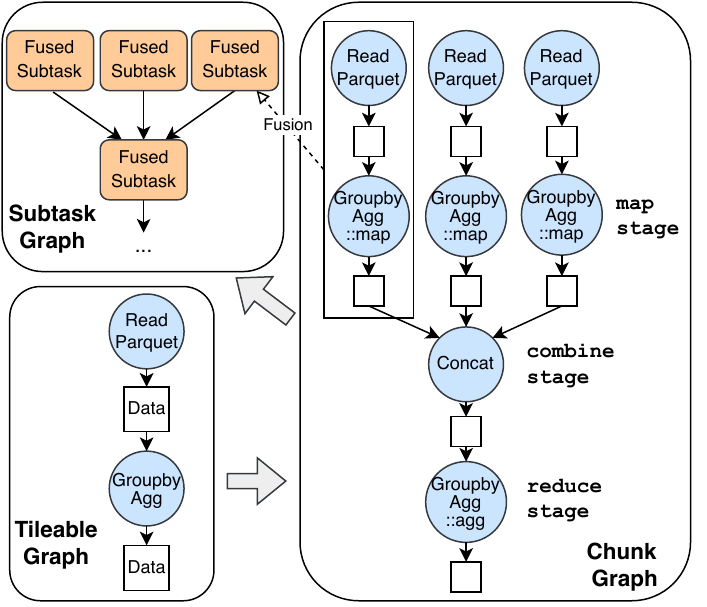}
    \caption{Dataframe example 1.}
    \label{fig:dataframe_graph}
  \end{subfigure}
  \begin{subfigure}{0.3\textwidth}
    \centering
    \includegraphics[width=0.98\linewidth]{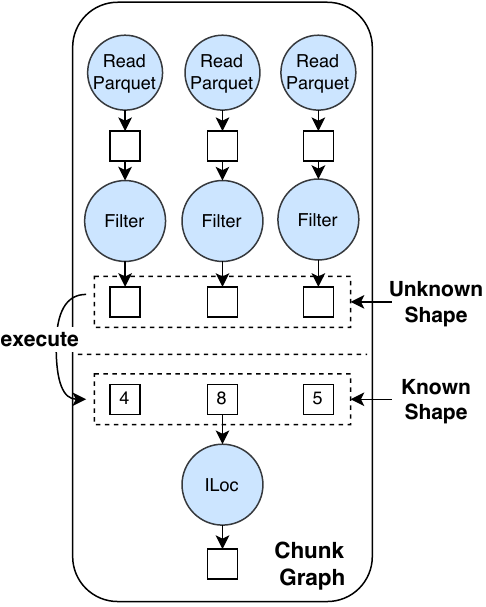}
    \caption{Dataframe example 2.}
    \label{fig:iterative_tiling}
  \end{subfigure}
  \caption{Illustration of computation graphs.}
  \label{fig:computation_graph}
  \vspace{-0.2in}
\end{figure*}

Figure~\ref{fig:workflow} depicts the workflow of the three types of graphs. Figure~\ref{fig:computation_graph} illustrates the computation graph for the three examples provided in Listing~\ref{lst:api_example}.
More specifically, Figure~\ref{fig:computation_graph}~(a) represents a fragment of the chunk graph for the QR decomposition example, while Figure~\ref{fig:computation_graph}~(b) and~(c) depict the pipelines for the dataframe examples. 
Note that Figure~\ref{fig:computation_graph} is only for illustration purposes, that the real graphs may have far more nodes than what we show here. 

Generally, each API offered by Xorbits is internally defined as an operator. In our computation graphs, operators are symbolized as circles, while squares signify data placeholders.
In this case, the \texttt{xorbits.pandas.read\_parquet} operation is implemented as the \texttt{ReadParquet} operator. For each operator, Xorbits implements three core methods: \texttt{\_\_call\_\_}, \texttt{tile}, and \texttt{execute}. These three methods align with distinct aspects of the processing pipeline: the \texttt{\_\_call\_\_} method corresponds to the tileable graph, the \texttt{tile} method associates with the chunk graph, and the \texttt{execute} method relates to the subtask graph.

\textbf{Tileable Graph}. 
The tileable graph represents a high-level, coarse-grained structure functioning as a logical plan. Xorbits calls the \texttt{\_\_call\_\_} method of each operator on the client and translates the user code into a node within the tileable graph. Note that at this stage, the tileable graph has not yet been divided into multiple partitions or chunks.

\textbf{Chunk Graph}. 
The tileable graph is then submitted to the supervisor, where every operator's \texttt{tile} method will be invoked.
The \texttt{tile} method divides the data into multiple chunks according to the data size or other relevant cues.
For complex operators like \texttt{groupby.agg}, the \texttt{tile} method adds internal nodes (i.e., the GroupbyAgg::map, Concat, and GroupbyAgg:agg nodes in Figure~\ref{fig:computation_graph} corresponding to the \textit{map}, \textit{combine}, and \textit{reduce} stages, respectively). The \textit{map} stage ingests the upstream chunk data and produces intermediate key-value pairs. The \textit{combine} stage is a pre-aggregation phase that combines a subset of chunks. 
In contrast to the MapReduce programming model~\cite{dean2004mapreduce} that uses persistent storage to store the key-value pairs, Xorbits is an in-memory computing engine. All intermediate results are retained within our dedicated storage service (Section~\ref{subsec:storage}). We add the \textit{combine} stage to avoid too many chunks aggregating into a single worker node, which may overwhelm the worker's memory. Our \textit{combine} stage's pre-aggregating can also improve performance. The \textit{reduce} stage will aggregate and convert the key-value pairs into the actual result. Note that each operator has its unique semantics, and not all operators would use the aforementioned \textit{map-combine-reduce} stages.

\textbf{Subtask Graph}. 
The subtask graph, or the coarse-grained physical execution plan, is optimized from the chunk graph. Although the chunk graph and the subtask graph resembles each other, there are two differences. 1) Nearby nodes in the chunk graph are fused to a subgraph called subtask. 2) Every subtask in the graph is assigned with scheduling information indicating which worker it should run on.
In Figure~\ref{fig:computation_graph}~(b), the \texttt{ReadParquet} and the \texttt{GroupbyAgg::map} are fused to form a subgraph, which will be scheduled to a particular worker.
During the execution phase, the \texttt{execute} method is called on the workers. Single-node packages are the backends for calculation given the split chunk (i.e., pandas is the backend for dataframes, and NumPy for arrays). There is also ongoing backend development of CuPy~\cite{CuPy} and cuDF~\cite{RAPIDS} for GPU support.

\begin{figure}[ht]
    \centering
    \includegraphics[width=0.78\linewidth]{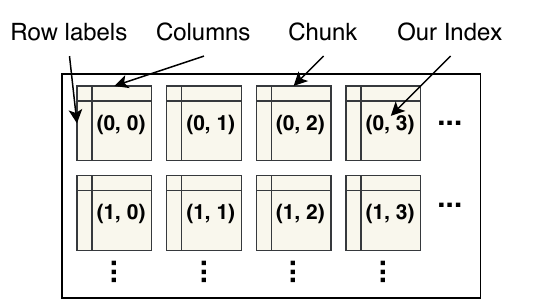}
    \caption{Distributed index.}
    \label{fig:distributed_index}
    \vspace{-0.2in}
\end{figure}

\textbf{Chunk}. Within the computation graph, circles represent operators, while squares symbolize chunks. Chunks serve as data placeholders, functioning as both the output for predecessor operators and the input for successor operators. An operator typically generates one or more chunks. As illustrated in Figure~\ref{fig:computation_graph}~(a), the \texttt{TensorQR} operator yields two output chunks: Q and R. Partitioning the entire dataset into chunks, whether it's done row-wise, column-wise, or cell-wise, depends on the semantics of the operator and the size of the data itself.

\textbf{Indexing and Ordering}.
One notable distinction between pandas dataframes and relational databases is the presence of row labels (or index) in pandas, often used for ordering-based operators like \texttt{iloc}. To preserve the semantics of pandas' index and multi-level index, we introduce a distributed index in each chunk. Illustrated in Figure~\ref{fig:distributed_index}, each chunk is a pandas dataframe and our distributed index consists of a two-value tuple $(r, c)$, where the $r$ indicates the vertical position of the chunk in the complete dataframe, while $c$ denotes the horizontal position. This distributed index enables Xorbits to locate any item in the original data, to implement operators like \texttt{iloc}, \texttt{transpose}, and to speedup lookups. 

\section{Dynamic Tiling}
\label{sec:dynamic_tiling}

After getting an overview of Xorbits in Section~\ref{sec:dynamic_tiling}, we present the dynamic tiling approach in this section. We first discuss when and why dynamic tiling is necessary (Section~\ref{subsec:necessity_dynamic_tiling}). Then we introduce the mechanism and implementation of dynamic tiling (Section~\ref{subsec:dynamic_tiling}). Finally, we showcase three typical scenarios that can benefit from dynamic tiling (Section~\ref{subsec:use_cases_dynamic_tiling}).

\subsection{Necessity of Dynamic Tiling}
\label{subsec:necessity_dynamic_tiling}

Performance and availability depend heavily on the tiling strategy because tiling too many chunks would increase overheads, while too few may cause memory overflow. 
On the other hand, the single-node packages do not need tiling, and these packages have no chunk- or partition-related parameters. Making users explicitly specify the tiling-related parameters would break the API compatibility and require expertise in parallel programming. 

\textbf{Unknown Shape Problem}.
Regarding the output data shape, there are mainly two types of operators: static and non-static. 
Static operators are those where the output data shape can be computed based on the input shape. A classic example is matrix multiplication.
Non-static operators are characterized by output data sizes that cannot be determined solely from the input shape. These operators' output sizes also depend on the data content.
Hence, the outputs' shapes of these operators are unknown, making tiling the rest of the data science pipeline notably difficult, as the exact shapes remain unknown until execution.
Non-static examples include \texttt{df["col"] < 1}, \texttt{groupby}, \texttt{merge}, \texttt{drop\_duplicates}, etc.
A pipeline with only static operators is easy to tile because the output of each operator's shape is fixed, given the shape of the data source. However, it is challenging to tile pipelines containing numerous non-static operators before execution. Because the intermediate results' shapes differ significantly from the original data sources, precisely determining the partitions of every operator is difficult.

\subsection{Dynamic Tiling}
\label{subsec:dynamic_tiling}

\begin{figure}
  \centering
  \begin{subfigure}[b]{0.4\textwidth}
    \includegraphics[width=0.9\textwidth]{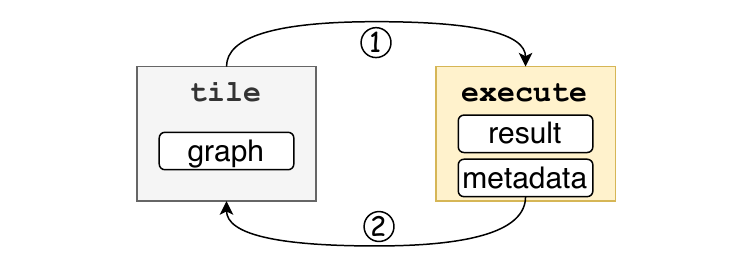}
    \caption{Switching between tiling and execution.}
    \label{fig:tile_execute}
  \end{subfigure}
  \begin{subfigure}[b]{0.5\textwidth}
    \includegraphics[width=\textwidth]{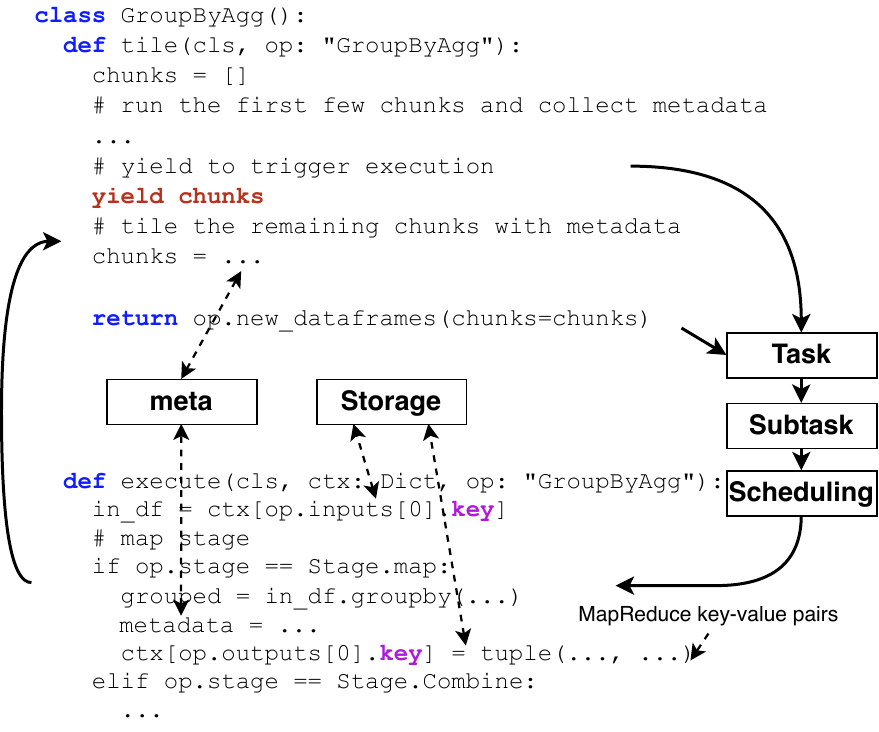}
    \caption{An example of operator implementation.}
    \label{fig:op_example}
  \end{subfigure}
  \caption{Dynamic tiling.}
  \label{fig:dynamic_tiling}
  \vspace{-0.25in}
\end{figure}

Our dynamic tiling approach leverages the metadata collected from execution to tile during the graph construction phase. Figure~\ref{fig:dynamic_tiling}~(a) shows how this procedure works. Xorbits begins by creating a coarse-grained chunk graph (Step \raisebox{.5pt}{\textcircled{\raisebox{-.9pt} {1}}}) and running the operator on the first few chunks. Xorbits would get the metadata (e.g., shape, columns, dtype, etc.) and then store it in the meta service (Step \raisebox{.5pt}{\textcircled{\raisebox{-.9pt} {2}}}) so that the tiling process can later access it. This type of metadata enables Xorbits to create an optimized computation graph.

\textbf{Yield}. The core technology underlying Xorbits' dynamic tiling is the ability to switch between tiling (i.e., the graph construction phase) and execution seamlessly. Xorbits generates the computation graphs on the fly by leveraging Python's \texttt{yield} mechanism. Figure~\ref{fig:dynamic_tiling}~(b) shows a code snippet when implementing the \texttt{GroupByAgg} operator. In the \texttt{tile} method, metadata (in this case, the actual data size after aggregation) is needed but currently missing, Xorbits will \texttt{yield} to trigger execution. Unlike the \texttt{return} keyword, which terminates the execution of the function, \texttt{yield} returns and pauses at where it is called. After the first few chunks are executed, and the metadata is collected, the function resumes from the same point. 
With the collected metadata, Xorbits can find the optimal way to tile the remaining chunks efficiently, and the optimized chunk graph will be scheduled again to workers. Although the \texttt{yield} solution sounds simple, developing the whole system is a non-trivial work, as Xorbits provides the services (e.g., task, subtask, scheduling, meta, storage, etc.) to guarantee transition between tiling and execution. 

\textbf{Iterative Tiling}.
As data science workloads usually contain a series of operators whose outputs' shapes are unknown, the dynamic tiling is done iteratively. Xorbits iterates over each operator and switches between tiling and execution if the metadata of that operator is needed. When encountering an operator that lacks metadata, Xorbits interrupts the tiling process, pauses, and submits a partial computation graph for execution. After the metadata (e.g., shape, columns, dtype, etc.) is updated, Xorbits continues the tiling process. In the second dataframe example of Listing~\ref{lst:api_example}, the output shape of \texttt{df["col"] < 1} is unknown until execution. It is impossible to know which chunk contains the tenth row of the filtered dataframe. This is where iterative tiling is utilized. The chunks corresponding to \texttt{filtered} are submitted for execution, and Xorbits will update the chunk shape based on the result. As shown in Figure~\ref{fig:computation_graph}~(c), suppose the initial dataframe \texttt{df} is divided into three chunks, and after filtering, the lengths of each chunk are 4, 8, and 5, respectively. To get the tenth row, we only need to append an \texttt{ILoc} operator to the second chunk to obtain the final result.

\subsection{Use Cases for Dynamic Tiling}
\label{subsec:use_cases_dynamic_tiling}

Dynamic tiling can be applied to many operators and scenarios. Here, we showcase three examples to illustrate how it helps to accelerate workloads and prevent OOM issues.

\begin{figure}
  \begin{subfigure}{0.3\textwidth}
    \centering
    \includegraphics[width=1\linewidth]{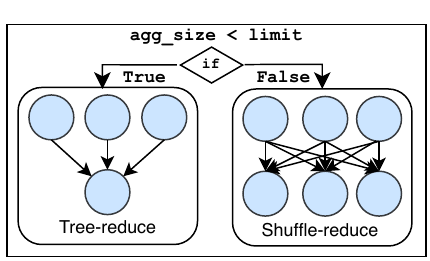}
    \caption{Auto Reduce Selection.}
    \label{fig:auto_reduce}
  \end{subfigure}
  \begin{subfigure}{0.18\textwidth}
    \centering
    \includegraphics[width=0.97\linewidth]{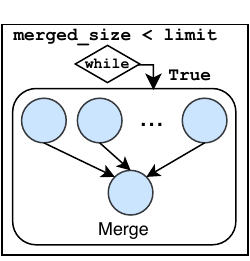}
    \caption{Auto Merge.}
    \label{fig:auto_merge}
  \end{subfigure}
  \caption{Typical use cases for dynamic tiling.}
  \label{fig:dynamic_tiling_cases}
  \vspace{-0.25in}
\end{figure}

\textbf{Auto Reduce Selection}.
One illustrative example is how to choose the optimal \textit{reduce} algorithm automatically. Figure~\ref{fig:dynamic_tiling_cases}~(a) shows the \textit{tree-reduce} and \textit{shuffle-reduce} algorithms widely used for operators like \texttt{groupby}. 
\textit{Shuffle-reduce} introduces communication overhead as it dispatches data to all downstream reducers, whereas tree-reduce transmits data solely to the combined nodes.
While \textit{tree-reduce} offers speed and simplicity, it is only efficient when the aggregated data is small. 
\textit{Tree-reduce} may encounter memory overflow as the data volume grows.
Therefore, a trade-off exists between performance and availability.
Other systems choose the reduce algorithm according to rules or manually specified by users. 
Given that most users lack enough knowledge of the \textit{reduce} mechanism, manual configurations by users could lead to memory issues or performance degradation.
With the dynamic tiling technique, Xorbits can intelligently choose the optimal \textit{reduce} algorithm based on the metadata collected during execution.
To illustrate, let's consider the \texttt{groupby.agg} operation. 
Initially, Xorbits builds a temporary chunk graph and runs on the first few chunks, obtaining the aggregated and raw input data sizes. This metadata is then added to the meta service and can subsequently be applied in tiling the remaining chunks. \
If the size of the aggregated data falls below a predefined threshold, Xorbits opts for the \textit{tree-reduce} structure; otherwise, it selects the \textit{shuffle-reduce}. Importantly, this entire process runs seamlessly without requiring user intervention.

\textbf{Auto Merge}. Large graphs would lead to the overhead of graph dispatching and graph execution.
Our auto merge mechanism can prevent this problem. 
In Xorbits, the configuration file predefines a chunk size limit, which serves as an upper bound for data chunk tiling.
Initially, we may get a large chunk graph with numerous small chunks, potentially causing a substantial performance bottleneck.
To keep the graph small and simple, Xorbits merges chunks in the \textit{combine} stage. Given the metadata (in this case, the chunk size) collected from the execution phase, Xorbits keeps concatenating data chunks until the merged chunks reach the predefined size limit.
This process is illustrated in Figure~\ref{fig:dynamic_tiling_cases}~(b).

\textbf{Deferred Evaluation}. In contrast to lazy systems where users are required to explicitly trigger execution, Xorbits seamlessly merges both lazy and eager modes.
We term this approach ``deferred evaluation," signifying a delayed evaluation until results are needed, without mandating users to trigger computation.
Since most single-node libraries only support eager mode, this feature ensures that Xorbits is compatible with those libraries and more user-friendly for exploratory tasks within Jupyter Notebooks.
In such environments, users often rely on immediate feedback for their subsequent actions.
The underlying technology for deferred evaluation is straightforward, facilitated by Xorbits' ability to seamlessly transition between graph tiling and execution.
We achieve this by customizing the \texttt{\_\_repr\_\_} method within our \texttt{Tensor} and \texttt{DataFrame} classes, where programs will invoke the \texttt{execute} function to activate the evaluation.
When users need to materialize the results, e.g., \texttt{print}, executions are launched, but users are unaware of it. 
Alternatively, users have the option to retrieve results without any delay by explicitly invoking \texttt{xorbits.run()}.
\section{Optimizations and Implementation Highlights}
\label{sec:implementation}

This section describes optimizations and key implementation highlights that underlie Xorbits' high-performance and scalable attributes.

\subsection{Data Science Graph Optimization}
\label{subsec:graph_optimization}

When generating the chunk graph and the subtask graph, Xorbits' optimizer conducts a series of optimizations to achieve better performance.

\begin{figure}[!ht]
    \centering
    \includegraphics[width=0.95\linewidth]{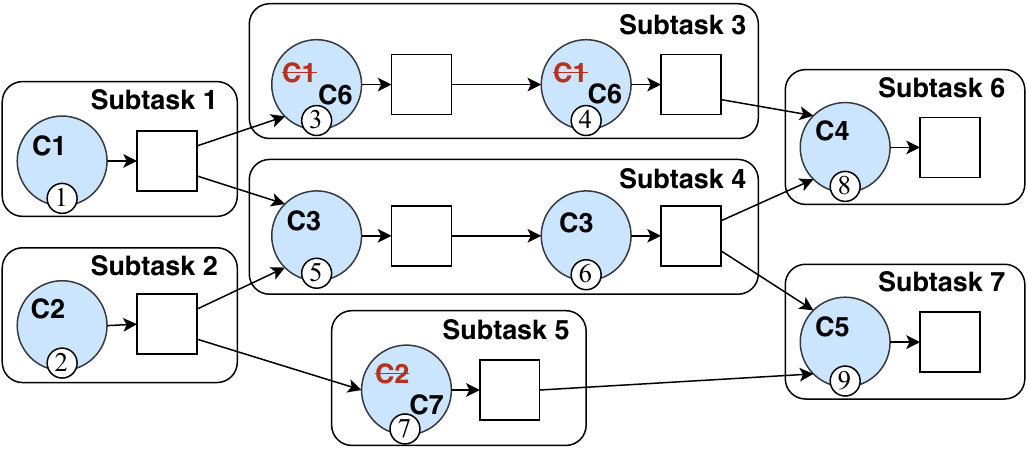}
    \caption{Coloring algorithm for graph-level fusion.}
    \label{fig:coloring}
    \vspace{-0.1in}
\end{figure}

\textbf{Graph-level Fusion}. Xorbits introduces a graph-level fusion algorithm based on coloring, to merge adjacent nodes.
A naive approach is to merge nodes straight in line, which is insufficient in our data science scenario as many other nodes are not involved in the fusion process.
The graph-level fusion algorithm, based on coloring, assigns distinctive colors to each node in the chunk graph. Nodes sharing the same color are candidates for merging into a subtask.
This process is illustrated in Figure~\ref{fig:coloring}, where the C label following numerical values represent specific colors, as seen with "C1" signifying Color 1. The numbers within the circles serve to differentiate between various operators; for instance, \raisebox{.5pt}{\textcircled{\raisebox{-.9pt} {1}}} represents Operator 1.
The coloring process consists of three main steps. In the first step, initial nodes in the graph are assigned colors (C1 for Operator \raisebox{.5pt}{\textcircled{\raisebox{-.9pt} {1}}} and C2 for Operator \raisebox{.5pt}{\textcircled{\raisebox{-.9pt} {2}}}).
In the second step, colors are propagated based on the topological order. If a node has multiple predecessors, and all of these predecessors share the same color, the node inherits that color (e.g., C1 for Operator \raisebox{.5pt}{\textcircled{\raisebox{-.9pt} {3}}}). Otherwise, the node is assigned a new color (e.g., C3 for Operator \raisebox{.5pt}{\textcircled{\raisebox{-.9pt} {5}}}).
The third step involves reverse topological order propagation. In this step, each node is assessed alongside its successors in the forward topological order. If all of a node's successors possess colors different from the node itself, the node is skipped. However, if some successors share the same color while others have different colors, new colors are assigned to the successor nodes with the same color. For example, in this step, the color of Operator \raisebox{.5pt}{\textcircled{\raisebox{-.9pt} {3}}} is changed from C1 to C6, and the color of Operator \raisebox{.5pt}{\textcircled{\raisebox{-.9pt} {7}}} is changed from C2 to C7. These new colors propagate to the respective successors, such as C6 for Operator \raisebox{.5pt}{\textcircled{\raisebox{-.9pt} {4}}}.
After these three steps, all the nodes within the chunk graph are assigned a color label, and nearby nodes that share the same color label are merged into a subtask. The first two steps of the coloring algorithm identify nodes that are in a straight line, while the third step tries to find nodes that require separation. In this specific case, Operator \raisebox{.5pt}{\textcircled{\raisebox{-.9pt} {1}}} should not be combined with either Operator \raisebox{.5pt}{\textcircled{\raisebox{-.9pt} {3}}} or Operator \raisebox{.5pt}{\textcircled{\raisebox{-.9pt} {5}}}.

\textbf{Operator-level Fusion}. Operator-level fusion is carried out after the completion of graph-level fusion. In this process, we harness state-of-the-art operator fusion libraries, such as numexpr~\cite{mcleod2018Pydata} and JAX~\cite{bradbury2018JAX}. These packages achieve high performance by combining multiple operators into a single one, thereby preventing memory allocation for intermediate results and minimizing memory access. Before execution, Xorbits traverses the subtask graph, identifies operators suitable to fuse, substitutes these operators with a fused one, and subsequently evaluates the computation with the fused operator.

\textbf{Column Pruning}. 
Xorbits facilitates column pruning, akin to predicate pushdown~\cite{ullman1990Principles, yan2023Predicatea}, which removes unnecessary data before it is loaded into memory or transmitted over the network.
Xorbits registers the required columns into every operator when building the tileable graph. 
During the optimization phase, Xorbits traverses backward from the data sink, recording the columns needed for each operator. 

\subsection{Scheduling Data Science Subtasks}
\label{subsec:scheduling}

The purpose of graph optimization is to schedule subtasks better. In Xorbits, a ``band" is the basic unit for subtask scheduling and execution. A band can be a NUMA (Non-Uniform Memory Access) node or a GPU device. Every band associated a hostname (or IP address) and a specific computing device, namely a NUMA node or a GPU device. Every subtask should be assigned to a particular band. We employ a combination of breadth-first and locality-aware strategies to guarantee that neighboring subtasks remain contiguous and to enhance the overall scheduling efficiency.

\textbf{Breadth-first}. Xorbits first assigns the initial subtasks of the computation graph with the breadth-first strategy. Breadth-first means that Xorbits would find the initial subtasks that do not have predecessors and try to assign more initial subtasks to one worker. In Figure~\ref{fig:coloring}, Subtask 1 and Subtask 2 are the initial nodes. Xorbits would first assign Subtask 1 to the first worker. Subsequently, Xorbits continues to allocate new subtasks to this worker until no bands remain available.

\textbf{Locality-aware}. To minimize the data transfer overhead of the non-initial subtasks, Xorbits implements a locality-aware strategy. This implies that successor subtasks should ideally be scheduled on the same computing worker as their predecessors, thus reducing data transfer overhead. As depicted in Figure~\ref{fig:coloring}, it is far more efficient to assign Subtask 2 and Subtask 5 to the same worker.

\vspace{-0.02in}

\subsection{Storage Service of Intermediate Results}
\label{subsec:storage}

Even though CPU performance has increased dramatically in recent years, the storage layer is often the bottleneck due to storage devices' latency. When developing Xorbits, we find that the distributed storage service affects performance and scalability when data size grows. Consequently, building a distributed storage layer is of great importance. 
To meet the performance requirements, we design a storage service to hold the intermediate results of all the calculations (i.e., the chunks produced by all the operators).  
Note that we here are not discussing the persistent storage layer that holds the data source or data sink. 

\textbf{Storage Backend}. Xorbits presently provides multiple storage backends (e.g., shared memory, mmap, cuda, Vineyard~\cite{yu2023Vineyard}, Alluxio~\cite{li2014Tachyon}, etc.) that serve as the underlying infrastructure for the storage service. The implementation of the storage backend follows three key considerations. 
First, the storage backend must utilize memory hierarchy. We define several \texttt{StorageLevel}s, including memory, GPU, disk, and remote distributed filesystem. For example, the shared memory's \texttt{StorageLevel} is memory. Users can either only use main memory, or combine main memory with disk and spill data to disk when data is large. Moreover, if the intermediate data is larger than the aggregated memory of a cluster, users can switch to a remote filesystem like Alluxio. 
Second, we must minimize data transfer. On each worker, Xorbits starts with \texttt{multiprocessing} module, and the data transfer between processes would introduce overhead. We adopt pickle5~\cite{pickle5} to achieve zero-copy data access between processes. We also add Vineyard support for data sharing between different data systems, which can reduce (de)serialization overheads.
Third, the storage backend is a layer of abstraction that hides the data access operations. Xorbits uses a unique ID (the \texttt{key} highlighted in Figure~\ref{fig:dynamic_tiling})~(b) for data indexing. Each storage backend offers \texttt{put} and \texttt{get} methods, with \texttt{key} as one parameter, to read and write data.
In this way, each worker can read and write data by indexing the \texttt{key} without knowing where the data actually is.

\textbf{Shuffling}. Using the abstraction provided by the storage service, Xorbits implements shuffling by writing data chunks to the storage service. Each subtask is scheduled to a band, and every data chunk has a distinct \texttt{key}. Xorbits maintains a dictionary that tracks which band each data chunk is on. Slightly different from non-shuffle data accesses, the shuffling data needs to be sent to the specified bands. We also optimize the shuffling by aggregating all the shuffling data together to reduce data transfer overheads. 

\subsection{Auto Rechunk}
\label{subsec:auto-rechunking}

When the underlying operator requires particular input sizes and shapes, our auto rechunk mechanism automatically adapts chunk sizes to fulfill these requirements. This eliminates the need for users to manually specify chunks or partitions, thereby preserving the compatibility of Xorbits' APIs with the original single-node packages.

\begin{figure}[t]
\begin{algorithm}[H]
  \caption{Auto Rechunk}
  \label{alg:auto_rechunking}
  \small
  \begin{algorithmic}[1]
    \algrenewcommand\textproc{}
    \Function{auto\_rechunk}{$shape, dim\_to\_size, itemsize, \newline \hspace*{1.6em} config$} 
    \State $max\_chunk\_size \gets config.chunk\_limit$

    \For{$i < len(shape)$}
        \If{$i$ not in $dim\_to\_size$}
        \State $left\_dim\_to\_size[i] \gets $ empty list
        \State $left\_unsplit[i] \gets shape[i]$
        \EndIf
    \EndFor
    \While{$True$}
        \State $nbytes \gets $ all\_items in $dim\_to\_size \times itemsize$ 
        \State $divided = max\_chunk\_size \div nbytes$
        \State $left\_dims \gets len(left\_dim\_to\_size)$
        \State $cur\_size = max(divided^{\frac{1}{left\_dims}}, 1)$
        \For{$j, ns$ in $left\_dim\_to\_size$}
            \State $unsplit \gets left\_unsplit[j]$
            \State $ns \gets$ \textbf{concat} $min(unsplit, cur\_size)$
            \State $left\_unsplit[j] \gets left\_unsplit[j] - ns[-1]$
            \If{$left\_unsplit[j] \leq 0$}
                \State $dim\_to\_size[j] \gets ns$
                \State $left\_dim\_to\_size[i] \gets $ empty list
            \EndIf
        \EndFor
        \If{$len(left\_dim\_to\_size) = 0$}
           \textbf{break}
        \EndIf
    \EndWhile
    \State \Return $dim\_to\_size$
    \EndFunction
  \end{algorithmic}
\end{algorithm}
\vspace{-0.3in}
\end{figure}

\textbf{Array Auto Rechunk}. For array operators like \texttt{qr} or \texttt{svd}, Xorbits uses Algorithm~\ref{alg:auto_rechunking} to choose the appropriate chunk size automatically. \texttt{shape} represents the raw data size before tiling. \texttt{dim\_to\_size} is a dictionary where the key is the dimension, and the value is the chunk size we want to partition on that dimension. $\{1: 10000\}$ indicates that there are 10,000 elements in the chunked data' second dimension (index begins with 0). \texttt{itemsize} is the number of bytes one array item occupies. The algorithm will return the right chunk size for each dimension. 
Take the \texttt{qr} operator for example. Both Xorbits and Dask adopt a MapReduce-based algorithm~\cite{dean2004mapreduce}. 
However, before invoking the MapReduce-based QR algorithm, Xorbits informs Algorithm~\ref{alg:auto_rechunking} that the chunked matrices are \textit{tall-and-skinny} via the \texttt{dim\_to\_size} parameter. This prevents users from manually selecting the appropriate chunk size. 
The auto rechunk algorithm returns the ideal chunk size given the input data shape. For instance, to adhere to the \textit{tall-and-skinny} rule, if the raw input shape of QR is $(10000, 10000)$, Xorbits specifies the \texttt{dim\_to\_size} with $\{1: 10000\}$. 
The chunk sizes determined by the auto rechunk algorithm are as follows: $(1677, 10000), (1677, 10000), ..., (1615, 10000)$.

\section{Evaluation}

In this section, we conduct experiments to evaluate the performance of Xorbits with different workloads.
Specifically, we seek to answer the following questions:

\begin{enumerate}
  \item What is the end-to-end performance of Xorbits compared with other frameworks, and how well does Xorbits scale data science workloads? 
  \item How much do our optimizations like dynamic tiling and graph fusion accelerate execution?
  \item How well can Xorbit cover APIs and use cases of single-node libraries?
\end{enumerate}

\subsection{Experiment Setup}
\label{sub:exp_setup}

The experiments have been carried out on the AWS r6i instance family. We start the supervisor of Xorbits (or the corresponding component of Spark, Dask, and Ray) on a r6i.large instance. All the workers are run on 16 r6i.8xlarge instances. Each of these instances has 32 vCPUs and 256GB memory. Different experiments use a subset of these instances or all of them. 

\textbf{Benchmarks}. We choose three distinct kinds of workloads: data science pipelines, ad-hoc queries, and array computing. Table~\ref{tab:benchmarks} shows the overview of all the workloads we use to benchmark different systems. 
For data science pipelines, we concentrate on the data preprocessing and feature engineering phases, which are common scenarios for pandas and NumPy.
We opt for a data science workload from TPCx-AI~\cite{brucke2023TPCxAI}, an industry standard, as well as those from Kaggle competitions (census and plasticc) that reflect real-world pipelines.
TPCx-AI benchmark comprises 10 use cases. We focus on use case (UC) 10 because other cases are too simple or complicated and require additional libraries.
For evaluating decision-making and analytical processing performance, we use TPC-H~\cite{boncz2014TPCH}. All 22 SQL queries are rewritten using the pandas API, with an emphasis on large datasets of SF100 and SF1000.
The array benchmark covers scientific computing workloads like linear regression (LR) and QR decomposition. These workloads span various scenarios, including DS preprocessing, machine learning (ML), and analytic processing (AP). In Table~\ref{tab:benchmarks}, we also show the number of workers we use when benchmarking different systems. We believe these workloads can effectively evaluate the scalability and compatibility of our Xorbits system. We run each workload seven times, excluding the maximum and minimum values, to obtain the average value.

\begin{table}[!ht]
\centering
\caption{Workloads to benchmark different systems.}
\label{tab:benchmarks}
\begin{tabular}{c|ccccc}
\hline
Workload                                                      & Size  & Format    & Workers & W/ IO & Type   \\ \hline
\begin{tabular}[c]{@{}c@{}}TPCx-AI \\ UC10 SF100\end{tabular} & 34GB  & CSV       & 2       & True       & DS, ML \\ \hline
census                                                        & 21GB  & CSV       & 1       & True       & DS, ML \\ \hline
plasticc                                                      & 20GB  & CSV       & 1       & True       & DS, ML \\ \hline
\begin{tabular}[c]{@{}c@{}}TPC-H\\ SF100\end{tabular}         & 36GB  & Parquet   & 4       & False      & AP     \\ \hline
\begin{tabular}[c]{@{}c@{}}TPC-H\\ SF1000\end{tabular}        & 358GB & Parquet   & 16      & False      & AP     \\ \hline
QR                                                            & Scale & Synthetic & 1-4     & True       & DS     \\ \hline
\begin{tabular}[c]{@{}c@{}}Linear\\ Regression\end{tabular}   & Scale & Synthetic & 1-4     & True       & DS, ML \\ \hline
\end{tabular}
\end{table}

\textbf{API Coverage}. In addition to these performance benchmarks, we add an API coverage benchmark to evaluate the compatibility of different systems. Most systems target the pandas' APIs, so we select 30 test cases from the airspeed velocity (asv) benchmark code of the pandas' GitHub repository. 
we primarily focus on \texttt{groupby}, \texttt{merge}, and \texttt{pivot}, because the statistics from the dataset~\cite{yan2020AutoSuggest}, which collected four million DS notebooks, indicate that these operators are the most popular ones.

\textbf{Baselines}. We compare Xorbits with pandas~\cite{mckinney2010Data}, pandas API on Spark~\cite{zaharia2016Apache}, Dask~\cite{rocklin2015Dask} DataFrame, and Modin~\cite{petersohn2021Flexible} on Ray~\cite{moritz2018Ray} for workloads related to dataframes. We compare Xorbits with Dask Array for the workloads based on arrays. We use the default configuration without any tuning or performance optimization for all of these systems. Table~\ref{tab:baseline_frameworks} shows all the frameworks and their versions we use. In the table, \emph{A} denotes array, and \emph{D} is short for dataframe.

\begin{table}[!ht]
\centering
\caption{Other data science frameworks used for baselines.}
\label{tab:baseline_frameworks}
\begin{tabular}{c|cccccc}
\hline
        & NumPy & pandas & Xorbits & PySpark & Dask   & Modin  \\ \hline
version & 1.26  & 2.1.1  & 0.6.3   & 3.5.0   & 2023.9 & 0.24.1 \\ \hline
API     & \emph{A}     & \emph{D}      & \emph{A} + \emph{D}   & \emph{D}       & \emph{A} + \emph{D}  & \emph{D}      \\ \hline
\end{tabular}
\end{table}
\vspace{-0.2in}

\subsection{DataFrame Performance}

\begin{figure*}
  \begin{subfigure}{0.3\textwidth}
    \centering
    \includegraphics[width=1\linewidth]{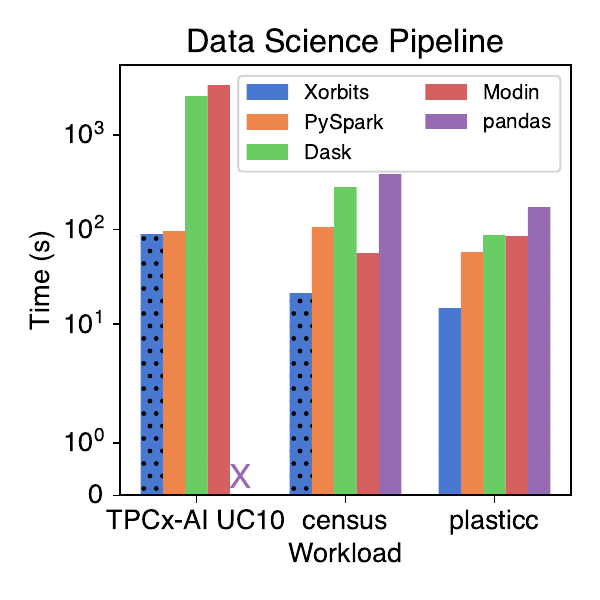}
    \caption{Data Science Pipeline.}
    \label{fig:data_science_pipelines}
  \end{subfigure}
  \begin{subfigure}{0.225\textwidth}
    \centering
    \includegraphics[width=1.\linewidth]{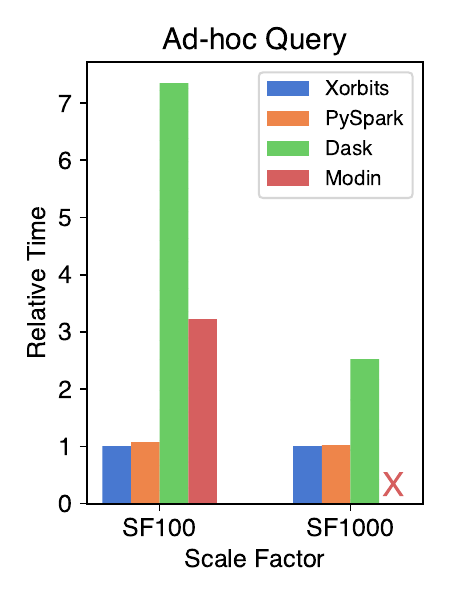}
    \caption{Ad-hoc Query.}
    \label{fig:tpch}
  \end{subfigure}
  \begin{subfigure}{0.2\textwidth}
    \centering
    \includegraphics[width=1.05\linewidth]{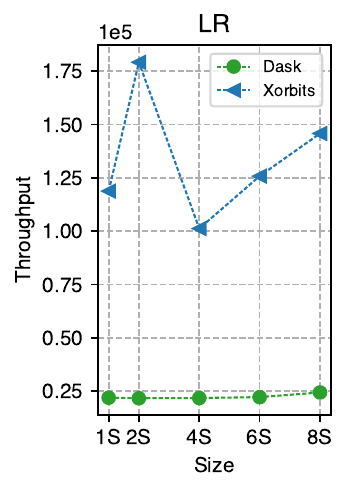}
    \caption{Linear Regression.}
    \label{fig:lr}
  \end{subfigure}
  \begin{subfigure}{0.2\textwidth}
    \centering
    \includegraphics[width=1\linewidth]{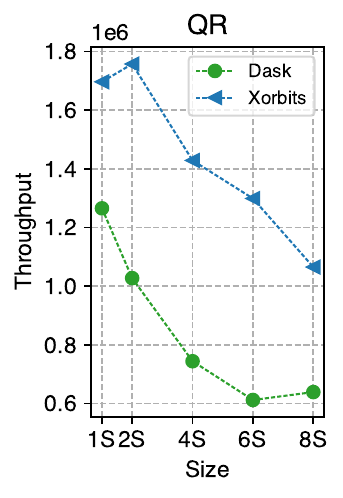}
    \caption{QR.}
    \label{fig:qr}
  \end{subfigure}
  \caption{End-to-end performance on different data science workloads.}
  \label{fig:end_to_end_performance}
    \vspace{-0.2in}
\end{figure*}

\textbf{Data Science Pipeline}.
Figure~\ref{fig:end_to_end_performance}~(a) shows the performance of data science pipelines. Overall, Xorbits outperforms the optimal baseline in each workload. These DS pipelines mainly contain operators like 
data filtering, missing data handling, and aggregated calculation for features. These operators are quite common, and all these frameworks can support them.
The TPCx-AI UC10-SF100 includes a customer file of 3.2MB and a financial transaction file of 34GB, which is much larger than the previous one. The pipeline joins the two imbalanced files based on customer IDs via the \texttt{merge} operator. There is a severe issue of data imbalance here. Both Dask and Modin cannot handle it well, and Xorbits is 29$\times$ and 37$\times$ faster than the two frameworks because they partition the task graph without knowing the actual data size. 
We observe the workers' CPU utilization; in this case, Dask and Modin can only utilize one CPU core, making the rest of the cores idle. This case illustrates that in scenarios involving data skew, simply partitioning the data during graph construction is insufficient. This is where our dynamic tiling approach excels, as our approach tiles the first chunks of the data, realizes that there is a data-skewing issue, and adjusts our computation graph accordingly.

The census and the plasticc datasets can fit into the memory of a single machine in our experimental environment.
Therefore, these two workloads show how these frameworks scale on a single machine to utilize all the CPU cores. Pandas is the slowest because it can run on only a single thread. 
Xorbits is 2.65$\times$ faster than Modin, which is the fastest on the census pipeline, and is 3.86$\times$ faster than PySpark on the plasticc pipeline. 

\textbf{Ad-hoc Query}.
Figure~\ref{fig:end_to_end_performance}~(b) demonstrates the performance of the ad-hoc queries on large datasets with the TPC-H benchmark,  with Xorbits standing out as the most compatible and fastest. We use two scale factors of TPC-H (SF100 and SF1000) to evaluate Xorbits' performance in analytic processing and to assess how our framework scales across nodes when dealing with large datasets. 
Table~\ref{tab:failed_queries} and Table~\ref{tab:failed_reason} provide clear evidence that many other systems frequently face challenges related to scalability and API compatibility.
For instance, three PySpark queries fail to run correctly when using code migrated from pandas, while Dask, PySpark, and Modin all encounter Out-Of-Memory (OOM) issues. 

The TPC-H benchmark is designed primarily to evaluate different SQL systems. It is more complex than typical data science workloads due to intensive operators, such as \texttt{groupby.agg} and \texttt{merge}.
Internally, Xorbits, Dask, and Modin use pandas as the execution backend to distribute tasks to multiple nodes. PySpark, on the other hand, translates pandas-like code into Spark's logical and physical plans. Using the TPC-H benchmark to evaluate our system is fundamentally unfair, as pandas has no performance advantage over SQL-based systems. 
Despite pandas-based systems having limitations in SQL queries, Xorbits still outperforms PySpark. Since not all of these baselines can successfully execute all 22 queries, we exclude the unsuccessful ones and calculate the overall relative time compared to Xorbits. 
We tried hard to execute Modin on Ray with SF1000, but the Ray workers are often dead because of memory overflows and disk space shortages. 

\subsection{Array Performance}

We perform a weak-scaling test to evaluate how Xorbits scales array workloads. We adjust the input size of each test case as computational resources expand, maintaining a consistent per-socket problem size. We use the linear regression and QR workloads: the linear regression is a classical ML workload, and the QR decomposition is a typical scientific computing operator upon which SVD can be constructed. Figure~\ref{fig:end_to_end_performance}~(c) and~(d) show the throughput, which is calculated via the problem size divided by time. Xorbits far surpasses Dask on both of the two workloads. First, on average, Xorbits outperforms Dask by factors of 5.88 and 1.74 on the two workloads. Second, when we raise the computing resources from one CPU socket to two, both workloads show a performance improvement. Because our machines have 2 CPU sockets, Xorbits can effectively schedule subtasks according to the NUMA sockets available and utilize the memory access patterns. Third, Xorbits shows higher throughput on the linear regression workloads as we increase the computing resources. Thus, Xorbits is a promising backend for a scalable ML toolkit because array APIs can also implement other ML algorithms. 
Fourth, Both Xorbits and Dask employ NumPy's \texttt{qr} as the backend, and the same MapReduce algorithm \cite{benson2013Direct} when implementing the distributed version of QR. Xorbits' auto rechunk mechanism not only partitions data more efficiently but also avoids manually rechunk operations.

\subsection{Ablation Study on Dynamic Tiling and Graph Optimization}

\begin{figure}
  \begin{subfigure}{0.23\textwidth}
    \centering
    \includegraphics[width=1\linewidth]{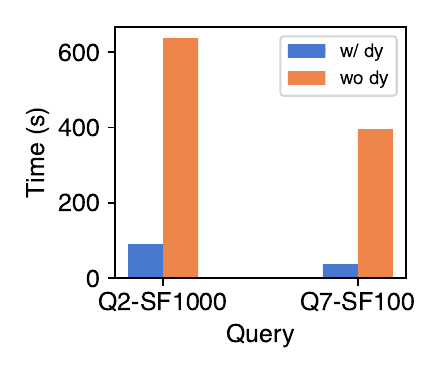}
    \caption{Dynamic Tiling.}
    \label{fig:dynamic_tiling_ab}
  \end{subfigure}
  \begin{subfigure}{0.23\textwidth}
    \centering
    \includegraphics[width=1.\linewidth]{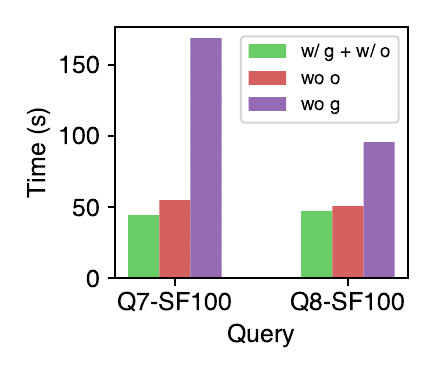}
    \caption{Graph Fusion.}
    \label{fig:fusion_ab}
  \end{subfigure}
  \caption{Ablation study of dynamic tiling and graph fusion. ``dy" means dynamic tiling, ``g" stands for graph-level fusion, and ``o" is for operator-level fusion.}
  \label{fig:ablation}
\end{figure}

We evaluate the potential for dynamic tiling and graph fusion to boost applications. For this purpose, we choose some TPC-H queries and alternately enable and disable these optimizations. The outcomes of these tests are illustrated in Figure~\ref{fig:ablation}.

\textbf{Dynamic Tiling}. Dynamic tiling can significantly speed up DS workloads, as depicted in Figure~\ref{fig:ablation}~(a) (``dy" denotes dynamic tiling). It is by default enabled in Xorbits for operations like \texttt{merge} and \texttt{groupby.agg}. In our experiments, Q2 has four \texttt{merge} operations, and Q7 has nine. When dynamic tiling is enabled, it yields 7.08$\times$ and 10.59$\times$ speed enhancement over its disabled state for these two queries, respectively. This outcome underscores the effectiveness of dynamic tiling.

\textbf{Graph Optimization}. 
Graph optimization, particularly the coloring-based graph-level fusion (``g" in Figure~\ref{fig:ablation}~(b)), plays a crucial role in enhancing performance. When the coloring-based graph-level fusion is activated, it results in a 3.80$\times$ and 2.04$\times$ acceleration in speed for Q7 and Q8, respectively, compared to when it is disabled. Operator-level fusion (``o" in Figure~\ref{fig:ablation}~(b)) is also preferred, as it can provide a 16\% improvement.

\subsection{API Coverage}

Table~\ref{tab:api_coverage} displays the coverage rate in the API coverage benchmark. Xorbits and Modin demonstrate superior compatibility with the original library, while Dask and PySpark encounter difficulties during code porting. 
In particular, the \texttt{merge} operators of Dask and PySpark do not support the sorting of join keys in the resulting dataframe. 
PySpark faces challenges with its aggregation functions; for instance, it is not user-friendly for user-defined aggregation functions, and does not support \texttt{NamedAgg}, which is the helper for column specific aggregation with control over output column names.
PySpark also offers a SQL-based interface\cite{sparkSQL}, which differs from conventional dataframe APIs. Online documentation and tutorials often mix SQL-based and pandas-based content, which increases the learning cost for users.
We believe this coverage benchmark can, to some extent, show the potential challenges users may encounter when transitioning from single-node libraries to these frameworks.

\begin{table}[!ht]
\centering
\caption{Coverage rate. Higher values are better.}
\label{tab:api_coverage}
\begin{tabular}{c|cccc}
\hline
        & Xorbits & Modin & Dask & PySpark \\ \hline
coverage rate & 96.7\%       & 96.7\%     & 46.7\%    & 36.7\%       \\ \hline
\end{tabular}
\vspace{-0.2in}
\end{table}

\subsection{Results and Findings}
\label{sub:results}

Among all the baseline frameworks, Xorbits stands out as the most scalable, high-performance, and compatible. 
It outperforms all other frameworks in data science, analysis and array workloads. 
While PySpark exhibits competitive performance, it encounters compatibility challenges. Users frequently have to rewrite their existing single-node code to accomplish specific tasks, posing a substantial burden.
Despite Modin's compatibility with pandas, it can only handle a moderate volume of data. Modin's lack of support for array operations restricts its capability for distributed ML tasks.
Dask, meanwhile, wrestles with both performance and compatibility issues.

\section{Related Work}
\label{sec:related}

There is a strong need to scale data science workloads horizontally~\cite{ gevay2021Efficient,fcs2023}, and numerous systems have attempted to do so by providing similar APIs of popular libraries of pandas~\cite{mckinney2010Data, mckinney2011Pandas} and NumPy~\cite{harris2020Array}. 

Apache Spark~\cite{zaharia2016Apache}, a popular big data engine, provides a Python interface called PySpark, enabling users to perform dataframe analysis on large datasets. 
However, PySpark users often encounter challenges related to API compatibility and are compelled to employ workarounds when migrating their code from pandas~\cite{fromToPySparkPandas, pysparkBestPractice}. 
While Spark provides an SQL optimization technique known as Adaptive Query Execution (AQE) that leverages runtime statistics for the selection of the execution plan~\cite{adaptiveQueryExecution, olma2020Adaptivea}, its performance on the PySpark DataFrame API appears to be insufficient. Spark runs on JVM, and JVM has limitations when integrating with the Python or C/C++ ecosystems and using GPU accelerators. PySpark lacks array APIs and is less interoperable than other Python-native libraries.
Dask~\cite{rocklin2015Dask} is another widely used Python library for parallel and distributed computing. At the low level, Dask designs a tasking mechanism. At the high level, it offers distributed arrays and dataframes. Dask's APIs are similar to those of NumPy and pandas, but Dask requires the explicit specification of chunks and partitions. 
Modin~\cite{petersohn2020Scalable} claims to be a scalable and drop-in
replacement for pandas. It formalizes the dataframe algebra and
outlines a set of decomposition rules. However, our empirical
study shows that it cannot handle data
skewing and fails on large datasets. Although it claims it can support array computing, it currently lacks NumPy-like APIs.
Ray~\cite{moritz2018Ray} and mpi4py~\cite{dalcin2005MPI} serve as general-purpose parallel computing engines, necessitating users to re-develop their single-node code entirely. 

Legate primarily concentrates on array computing. It is implemented with a runtime called Legion~\cite{bauer2012Legion} and offers a limited set of APIs compared to NumPy. JAX, on the other hand, provides interfaces similar to NumPy and utilizes XLA~\cite{sabne2020XLA} as its execution backend. While JAX, PyTorch~\cite{paszke2019PyTorch}, and TensorFlow~\cite{abadi2016TensorFlow} have gained widespread acceptance for deep learning training and inference, they may not be as well-suited for data science preprocessing.

Auto-Suggest~\cite{yan2020AutoSuggest} conducted research into the behavior of data scientists, collecting millions of data science notebooks. It subsequently introduced an automated method for generating data preparation code. MagicPush~\cite{yan2023Predicatea} implemented predicate pushdown techniques for data science pipelines.

\section{Conclusion}

Data scientists frequently perform various tasks on increasing volumes of data, typically employing tools like pandas and NumPy. It is of great significance to extend pandas and NumPy to adapt to modern hardwares. Existing distributed frameworks suffer from scalability and usability issues. They do not partition big data well by constructing the computation graph only before execution. Xorbits can scale well while providing compatible interfaces by designing three types of computation graphs and introducing a novel dynamic tiling approach. Xorbits' dynamic tiling can switch between graph building and graph execution and thus can tile data automatically by leveraging the execution metadata. Extensive experiments demonstrate that Xorbits significantly outperforms the state-of-the-art frameworks on various data science workloads.

\balance

\section*{Acknowledgment}

This work is partly funded by the China National Science Foundation (Grant No.62272466 and No.62322213). We thank AWS and PCC@RUC for providing computing resources. We also thank all the contributors of Xorbits and Mars.

\end{document}